# Magnetic ordering, Kondo effect and superconductivity in $Ce_{2-x}(La, Y)_xRhSi_3$

Kausik Sengupta, S. Rayaprol, and E.V. Sampathkumaran

Tata Institute of Fundamental Research, Homi Bhabha Road, Mumbai 400 005

*Abstract*

*The influence of positive and negative chemical pressure on the magnetic behavior of the compound, $Ce_2RhSi_3$, crystallizing in a $AlB_2$-derived hexagonal structure and ordering antiferromagnetically around ($T_N =$) 7 K, is investigated by studying electrical resistivity ($\rho$) and magnetic susceptibility ($\chi$) behavior of the solid solutions formed by gradual substitution of La/ Y for Ce. While typical features associated with a transformation from the Kondo-lattice to single-ion Kondo effect are seen in the low temperature $\rho$ data of both the solid solutions, there are profound differences in their behavior at higher temperatures. In particular, Y is apparently more effective in enhancing the Kondo temperature. $La_2RhSi_3$ is found to be superconducting below 3 K.*

## INTRODUCTION

Cerium-based intermetallic compounds continue to attract a lot of attention, primarily due to the competition between different interactions: crystal-field effects, magnetic order and the on-site Kondo-effect. In this regard, very little attention has been paid to the compound, $Ce_2RhSi_3$, crystallizing in an $AlB_2$-derived hexagonal structure [1] and found to order antiferromagnetically at 7 K [2]. It is of interest to explore the role of chemical pressure effects on the magnetism of this compound. With this primary motivation, we have initiated investigations on the solid solutions, $Ce_{2-x}R_xRhSi_3$ (R= La, Y; $0 \leq x \leq 2$). We report here the results of our initial investigations.

## EXPERIMENTAL DETAILS

Polycrystalline samples of $Ce_{2-x}R_xRhSi_3$ (R = La, Y; x = 0, 0.3, 0.5, 1, 1.5, 1.7, 2) were synthesized by arc-melting high purity constituent elements in an inert atmosphere. The molten ingots were annealed at 800° C for 5 days in vacuum and characterized by x-ray diffraction.. The dc electrical resistivity ($\rho$) behavior was obtained (1.5-300 K) using a standard four-probe method. Preliminary magnetization measurements were carried out employing a commercial magnetometer to augment our conclusions.

## RESULTS AND DISCUSSION

The lattice parameters a and c and unit cell volume (V) for the alloys $Ce_{2-x}La_xRhSi_3$ and $Ce_{2-x}Y_xRhSi_3$ are depicted in figure 1. The lattice parameters change linearly with x, indicating the validity of Vegard's law for both the solid solutions. It is obvious that, as expected, Y substitution compresses the lattice, whereas La substitution expands the lattice.

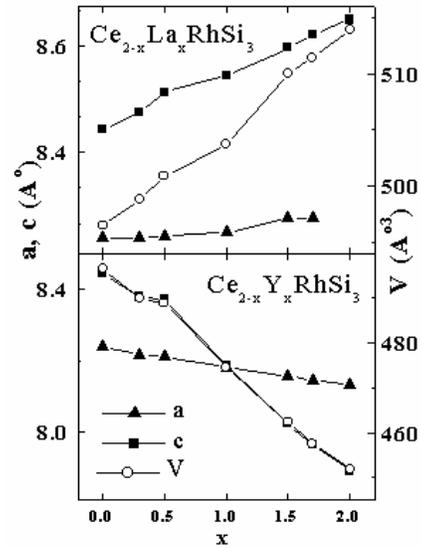

Fig. 1. Lattice parameters a, c and unit cell volume V.

The results of $\rho$ measurements are shown in Fig. 2 (normalized to respective 300 K values). For x= 0.0, $\rho$ is nearly constant above 150 K and there is a gradual fall below 150 K followed by an upturn below 15 K attributable to the Kondo effect. This kind of feature in $\rho$ is typical of interplay between the Kondo effect and the crystal-field effect [3]. With further lowering of T, there is a fall at 7 K arising from the onset of long-range antiferromagnetic order. As La/Y is substituted for Ce, the drop due to magnetic ordering gets gradually depressed towards lower temperatures and the upturn below 20 K gets more pronounced. For x>1.5, this upturn only could be seen without any drop and therefore $\rho(T)$ is dominated by features attributable to single-ion Kondo effect only above 1.5 K. Thus, Kondo lattice to Kondo-impurity transformation is brought out by these substitutions. We would like to stress on a difference on the $\rho$ behavior at high temperatures among these two solid solutions: There is a distinct broad maximum in the $\rho(T)$ plot around 150 K for x>0.5 for the Y series, which is absent for the La series. We believe that the origin of this feature for the Y series lies in the enhanced Kondo effect induced by positive chemical pressure induced by Y substitution, as it is a well-known fact that the strength of the Kondo interaction increases with V.

In order to render support to the above conjecture, we have carefully analyzed the magnetic susceptibility ($\chi$) data in the paramagnetic state (not shown here). The plot of inverse $\chi$ is found to be linear above 150 K for all values of x (<2). For x= 0.0, the negative sign with a large magnitude (-65 K) of the paramagnetic Curie temperature ($\theta_p$) compared to the value of $T_N$ indicates significant role of the Kondo effect. While the negative sign of $\theta_p$ is maintained for all Ce containing compositions, the magnitude decreases (increases) with increasing La (Y) concentration. Thus, for instance, for x= 1.0, in the case of La series, we get a value of about –35 K, whereas for the Y series, the corresponding value is about –180 K. These values clearly establish, from the knowledge of the direct relationship between $\theta_p$ and the Kondo temperature ($T_K$) in the field of the Kondo effect, that the Y substitution enhances $T_K$, whereas La does the reverse.

Another finding we have made in this investigation is that Y substitution is more effective in depressing $T_N$ than La substitution. Thus, for instance, for x= 0.5, in the case of La series, the obtained value (5.5 K) of $T_N$ nearly scales (even a bit higher!) with the concentration of Ce, whereas, in the case of Y series, one obtains a reduced value (4 K). This finding brings out the enhanced role of the Kondo effect (favoring non-magnetism) by Y substitution. This finding is in accordance with Doniach's model [4] to describe the competition between magnetism and the Kondo effect in Ce compounds.

Finally, the present $\rho$ data (as well as $\chi$, not shown here) clearly reveal that $La_2RhSi_3$ is superconducting below 3 K, uncommon finding among the ternary rare earths with this structure. In Y analogue, there is a resistive drop below 3 K, however without attaining zero value, implying that this compound may not be a bulk superconductor.

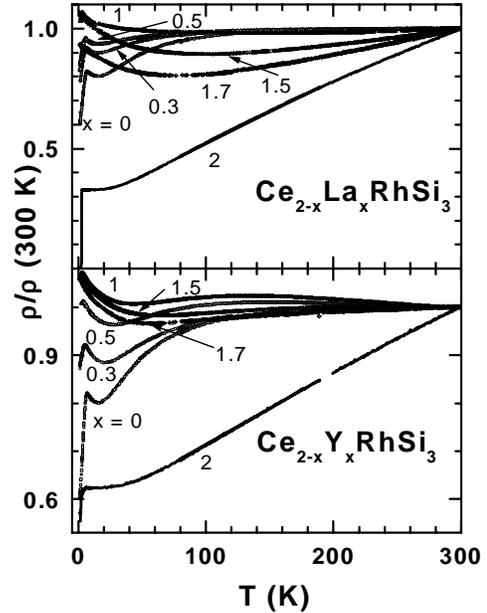

Fig. 2. Resistivity of $Ce_{2-x}R_xRhSi_3$ (R= La, Y).

## CONCLUSIONS
Kondo lattice to Kondo impurity transformation in $Ce_2RhSi_3$ is investigated by the substitution for Ce by La and Y. Y substitution is found to enhance the Kondo effect, resulting in pronounced differences in the $\rho$ behavior at higher temperatures (above 50K) while compared with the data for the La series. The present studies also show the existence of a superconducting transition for $La_2RhSi_3$ below 3 K.